\documentclass[preprint,authoryear,12pt,groupaddress]{elsarticle}

\usepackage{booktabs}
\usepackage{color}
\usepackage{lipsum}
\usepackage{float}
\usepackage[titletoc,toc,title]{appendix}
\usepackage{amssymb}
\usepackage{epsfig}
\usepackage{amsmath}
\usepackage{times}
\usepackage{subfigure}
\usepackage{graphicx}
\usepackage{booktabs}
\usepackage{colortbl}
\usepackage{natbib}
\usepackage{color}

\begin{document}

\title{The dose-dense principle in chemotherapy}

\author[label1]{\'{A}lvaro G. L\'{o}pez}
\author[label2]{Kelly C. Iarosz}
\author[label3]{Antonio M. Batista}
\author[label1]{Jes\'{u}s M. Seoane}
\author[label4]{Ricardo L. Viana}
\author[label1,label5]{Miguel A. F. Sanju\'an}

\address[label1]{Nonlinear Dynamics, Chaos and Complex Systems Group.\\Departamento de F\'isica, Universidad Rey Juan Carlos, Tulip\'an s/n, 28933 M\'ostoles, Madrid, Spain}

\address[label2]{Instituto de F\'{i}sica, Universidade de S\~ao Paulo, 05315-970 S\~ao Paulo, SP, Brazil}

\address[label3]{Departamento de Matem\'{a}tica e Estat\'{i}stica, Universidade Estadual de Ponta Grossa, 84030-900 Ponta Grossa, PR, Brazil}

\address[label4]{Departamento de F\'{i}sica, Universidade Federal do Paran\'{a}, 81531-990 Curitiba, PR, Brazil}

\address[label5]{Institute for Physical Science and Technology, University of Maryland, College Park, Maryland 20742, USA}

\date{\today}

\begin{abstract}
Chemotherapy is a cancer treatment modality that uses drugs to kill tumor cells. A typical chemotherapeutic protocol consists of several drugs delivered in cycles of three weeks. We present mathematical analyses demonstrating the existence of a maximum time between cycles of chemotherapy for a protocol to be effective. A mathematical equation is derived, which relates such a maximum time with the variables that govern the kinetics of the tumor and those characterizing the chemotherapeutic treatment. Our results suggest that there are compelling arguments supporting the use of dose-dense protocols. Finally, we discuss the limitations of these protocols and suggest an alternative.\\
\end{abstract}

\maketitle



\section{Introduction}\label{sec:intro}

In order to assess the benefits of the different combination chemotherapeutic protocols, clinical experience reveals that simple trial-and-error, in the absence of guiding principles, is a rather slow and inefficient process \citep{nsh}. To establish these guiding principles, hypotheses have to be accompanied by mathematical models \citep{kavg,kav}, in such a manner that empirical data allows their rigorous falsification. Log-kill models have provided important progress in chemotherapy along the last forty years \citep{lokil}, specially concerning haematological cancers. However, randomized trials carried out in recent decades \citep{dd1,dd2,dd3} have demonstrated that the periodicity of the cycles is very important, as well. One of the reasons that support this fact is that tumors grow between cycles of chemotherapy. Even worse, some of these proliferating cells could be resistant to further treatment. Another reason is that there is evidence suggesting that the rate of destruction by chemotherapy is proportional to the rate of growth of the same tumor in the absence of therapy \citep{nsh}. This statement is formally known as the Norton-Simon hypothesis. According to it, and because many solid tumors follow Gompertzian or sigmoidal growth \citep{gomp0,gomp}, bigger tumors are less susceptible to therapy. All these facts have led to the introduction of the concept of dose-dense protocols in chemotherapy. Dose-dense chemotherapy is based on the increase in the frequency of drug delivery to avoid regrowth between cycles and achieve the maximum cancer cell kill. It has been an important breakthrough in the evolution of chemotherapy for breast cancer and lymphoma \citep{hud2}.

However, dose-dense protocols of chemotherapy are still being debated \citep{notwo}, and no general consensus has been reached on their beneficial properties. Moreover, as far as the authors are concerned, the role of dose-density compared to dose-intensity has not been addressed in previous modeling efforts \citep{panad,pinho,mixed} on chemotherapy. The importance of the Norton-Simon hypothesis and how it affects the dose-dense principle has not been rigorously established neither. This novel features can be studied by introducing protocols of chemotherapy that are in closer resemblance to those used in the clinical practice. For this purpose, we devise a new one-dimensional map for chemotherapy from a well-established continuous mathematical model. The discrete model is used to demonstrate the existence of a maximum time between cycles of chemotherapy for a treatment to be able to reduce the tumor mass. The continuous mathematical model assumes the the sigmoidal growth of tumors, although other types of growth can be considered to obtain similar conclusions. Concerning chemotherapy, it is represented by means of the Exponential Kill Model \citep{gard}, which was developed in the last decade, and includes features that go beyond traditional log-kill models. We do not need to assume the Norton-Simon hypothesis to develop our ideas. Nevertheless, we show that such hypothesis enhances the forcefulness of our results. Before developing our ideas, it is convenient to put on a clear physical basis some concepts commonly used in the arrangement of chemotherapeutic protocols, which might be potentially confusing. 

\section{Protocols of chemotherapy}

Chemotherapy can be used in several ways \citep{cama}. Among its different uses, it can be administered concurrently with other treatments, such as radiation and surgery (adjuvant chemotherapy). It can also be administered in a long-term setting at low-doses, to a patient who has achieved a complete remission, with the intent of delaying the regrowth of residual tumor cells (maintenance chemotherapy). Or it can be delivered to prolong life, to a patient whose cure is not likely (palliative chemotherapy). In all these cases, chemotherapy is commonly delivered in periodic cycles of three weeks\footnote{Information about standard protocols of chemotherapy has been drawn from:~http://www.bccancer.bc.ca/health-professionals/professional-resources/chemotherapy-protocols.}, which seems to be the time required for the organism to recuperate from the toxic side-effects of the therapy (\emph{e.g.} to replenish cells originated in the bone marrow). Finally, a cycle of chemotherapy consists of several drugs, which are frequently administered intravenously through continuous infusion. These infusions can last from half an hour to several hours.  Therefore, four elementary variables associated to a chemotherapeutic protocol can be distinguished. During a cycle, for each drug, there is a variable representing the dose administered $D$, another that symbolizes the total duration of the infusion $t_{a}$ and a third variable that represents the rate of elimination of such drug from the bloodstream $k$. Finally, one more variable representing the time $T$ between the successive cycles of the treatment is necessary. 

\begin{figure}
\centering
\includegraphics[width=0.60\linewidth,height=0.36\linewidth]{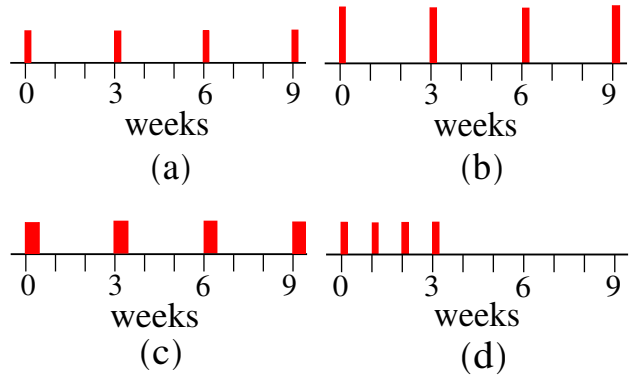}
\caption{\textbf{Protocols of chemotherapy}. (a) A reference protocol of chemotherapy consisting of four cycles of the same dose of drug, given every three weeks. (b) A protocol that is more dose-intense than the reference, because higher doses of the drug are administered (higher $D$). (c) A protocol that is more dose-intense, again because more drug is administered in a cycle, but through a longer continuous infusion (higher $t_{a}$). (d) A protocol consisting of four cycles, which is more dose-dense than the reference, because the frequency of the cycles is increased (lower $T$) to one week.}
\label{fig:1}
\end{figure}

There are two fundamental concepts related to these variables. The first is \emph{dose-intensity}, which is defined as the total dose of drug administrated during a treatment, divided by the duration of the treatment \citep{dd4}. The second concept is \emph{dose-density}, and it can be precisely defined as the period $T$ between the cycles of the treatment \citep{dd4}. Several protocols illustrating these two concepts are shown in Fig.~\ref{fig:1}. The confusing point is that, since dose-intensity is defined as an average, there are two ways in which it can be incremented. The first is to increase the dose of a drug (or the number of drugs) given in a cycle. The second is to shorten the treatment by an increase of dose-density (keeping fixed the total amount of drug delivered and the number of cycles). To avoid this ambiguity, dose-intensity must be defined at every instant of time, as the rate at which drug flows into the body $I(t)=d D/d t$. However, in practice, we also define the dose-intensity as an average. In particular, we say that the dose-intensity $\left< I \right>$ is the average of the instantaneous dose-intensity $I(t)$ over a cycle of chemotherapy. Mathematically, this can be written as
\begin{equation}
\left< I \right>=\dfrac{1}{T}\int_{0}^{T} I(t) d t.
\label{eq:1}
\end{equation}
Now, this concept is independent of dose-density and the only way to increase it is through an increase of dose or through the addition of more drugs to a cycle. To conclude, we recall that other concepts, such as the cumulative dose, do not play any role in our study, since drugs barely accumulate when the time between cycles is considerably longer than the half-lives of the drugs. For the same reason, the time along which drugs are administrated $t_{a}$ through continuous infusion, is neglected. 

\section{Model description}\label{sec:md}

To develop our ideas, a one-dimensional nonhomogeneous ODE model governing the dynamics of a solid tumor under the effect of cytotoxic drugs is considered. The kinetics of the tumor is assumed to be Gompertzian, although for mathematical simplicity, the logistic equation is used. The chemotherapeutic protocol is represented by means of the Exponential Kill Model, which was designed regarding \emph{in vitro} data \citep{gard} and has also been tested against \emph{in vivo} results \citep{validbul}. Therefore, the mathematical equation can be written as
\begin{equation}
\dfrac{d P}{d t}=r P(t) \left(1-\dfrac{P(t)}{K} \right)-b \left(1-e^{-\rho C(t)} \right) P(t), 
\label{eq:2}
\end{equation}
where $P(t)$ represents the tumor cell population, $r$ its maximum rate of growth and $K$ its carrying capacity. The second term represents the action of a cytotoxic agent, being $b$ the maximum fractional cell kill, $C(t)$ the concentration of the drug at the tumor site and $\rho$ the resistance of the tumor cells to such drug.

The effect of chemotherapeutic drugs is not exerted immediately, and there exist time delays imposed by their metabolism \citep{validbul}. However, these delays do not alter the results of this work, since they simply displace in time the treatment. Thus, concerning the pharmacokinetics, we simplify it as much as possible, following previous modeling efforts \citep{mixed}. This allows us to derive analytical results. Hence, we assume a one-compartment model and first order pharmacokinetics. The differential equation governing the concentration of the drug is
\begin{equation}
\dfrac{d C}{d t}=I(t)-k C(t), 
\label{eq:3}
\end{equation}
where $I(t)$ is the function representing the input of drug (the instantaneous dose-intensity) and $k$ is the rate of elimination of the drug  from the bloodstream, from which the half-life can be computed as $\tau_{1/2}=(\log_{e}2)/k$.
\begin{figure}
\centering
  \includegraphics[width=0.58\linewidth,height=0.33\linewidth]{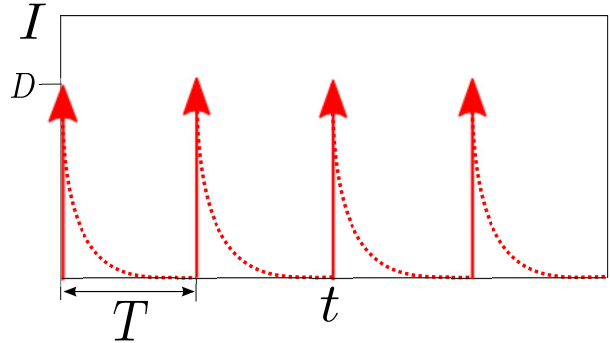}
\caption{\textbf{The drug input}. The arrows represent a protocol of four cycles of chemotherapy modeled through a series of Dirac delta functions. The dashed line represents the concentration of the drug as it is eliminated from the organism.}
\label{fig:2}
\end{figure}

If the drug is administered through an intravenous bolus or by a short continuous infusion, the time along which the drug is given (from minutes to a couple of hours) can be neglected compared to the time between cycles of chemotherapy (weeks), and we can safely approximate the input function as
\begin{equation}
I(t)=\sum_{m=0}^{N_{c}}{D \delta(t-m T)},
\label{eq:4}
\end{equation}
where $\delta(t)$ is the Dirac delta function and $T$ is the time elapsed between cycles. Therefore, every $T$ weeks a dose of drug $D$ is administered to the patient, during a treatment that comprises $N_{c}$ cycles of chemotherapy. Note that, with this approximation, the dose-intensity $\left< I \right>$, as computed from equation \eqref{eq:1}, is simply given by $D$. For the moment, we consider that the same dose of drug is administered with each cycle. In Sec.~\ref{sec:comb} our results are extended to more sophisticated protocols. For a representation of the protocol see Fig.~\ref{fig:2}. The solution to equation \eqref{eq:3} with a drug input given by equation \eqref{eq:4} at a time instant $t$ between the $(n-1)$-th and $n$-th cycle is
\begin{equation}
C(t)=D \dfrac{e^{n k T}-1}{e^{k T}-1}e^{-k t}.
\label{eq:5}
\end{equation}
If the drug is eliminated quickly (in comparison to the duration of the cycle) we can neglect the accumulation of the drug along the cycles of the treatment and use the approximation $C(t)=D e^{- k(t-(n-1)T)}$, with $t$ in the interval mentioned above. Or more simply
\begin{equation}
C(t)=D e^{-k t} (\text{mod}~T).
\label{eq:6}
\end{equation}

\begin{figure}
\centering
  \includegraphics[width=0.9\linewidth,height=0.48\linewidth]{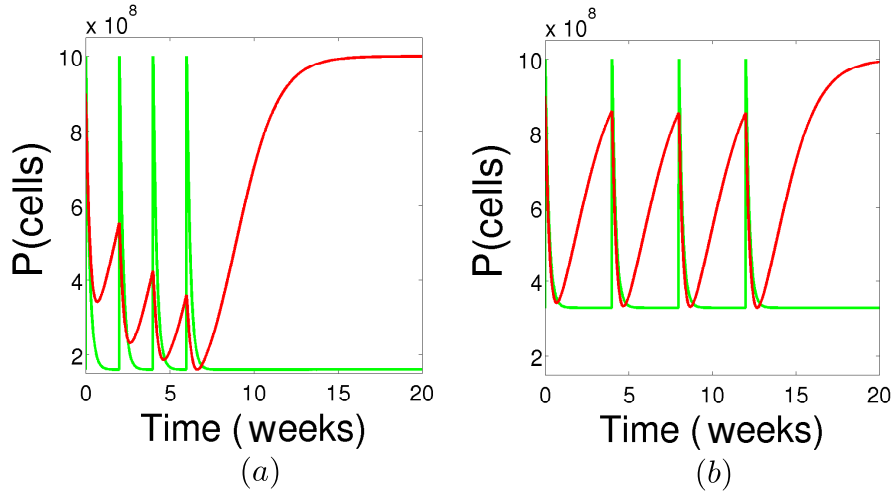}
\caption{\textbf{Two protocols of chemotherapy}. (a) A protocol consisting of two-week cycles ($T=2~ \text{weeks}$) is able to reduce the size of the tumor (red) progressively. (b) A protocol whose drugs are administered every four weeks ($T=4~\text{weeks}$) is insufficient to reduce the size of the tumor progressively. The time series of the drug concentration $C(t)$ is plotted for clarity (green), disregarding its specific values.}
\label{fig:3}
\end{figure}

We now briefly describe the main features of the model dynamics. To this end, we consider the following set of parameters, which are chosen in conformity with experimental data. The values of $r=0.8~\text{week}^{-1}$ and $K=1 \times 10^{9}$ cells have been borrowed from the literature \citep{validpilis}. Since the fastest growing tumor that can be imagined is an exponentially growing tumor with a constant rate value of $r=4.85~\text{week}^{-1}$, we are considering a quite aggressive tumor. The carrying capacity corresponds to a detectable tumor mass of approximately one gram. The values of $\rho=0.1~\text{mg}^{-1}$ and $b=2.8~\text{week}^{-1}$ are within values appearing in other work as well \citep{gard}. The dose of drug administered in these simulations is $D=60~\text{mg}$, while the rate of drug elimination $k=4.85~\text{week}^{-1}$ corresponds to a half-life of approximately one day. These are typical values of the drug doxorubicine, which is used in the treatment of locally advanced breast cancer, for example. Nevertheless, the effects of varying these parameters are inspected in Sec.~\ref{sec:sco}. As can be seen in Fig.~\ref{fig:3}, if the drug is effective destroying the tumor cell population (low resistance), the tumor is considerably reduced when the drug concentration is high. However, as soon as most of the drug has been eliminated, the tumor resumes its growth. Consequently, if the period of time between cycles is too long, the protocol leads to an oscillatory dynamics and cannot be effective, bearing in mind that typical half-life values of cytotoxic drugs span from several hours to several days. Unlike other cancer treatments, where the drugs are given for longer periods of time and the phenomenon of drug resistance is more pronounced \citep{hira}, in chemotherapy the resulting oscillations are in general undesired. An example is the use of adjuvant chemotherapy for the treatment of breast cancer, where the phenomenon of resistance has not been demonstrated to be more important than the effect of dose-density \citep{dd1}.

These facts suggests that, having fixed the remaining parameters of the model, there exists a maximum time between cycles of chemotherapy that permits a progressive reduction of the tumor. And then, obviously, protocols set at values of $T$ higher than such threshold (${i. e.}$, not enough dose-dense), are useless. A mathematical function that allows to estimate the relation between the threshold value of the cycle periodicity and the model parameters is derived in the following sections. 

\section{A one-dimensional map} \label{sec:1dm}

\begin{figure}
\centering
  \includegraphics[width=0.46\linewidth,height=0.42\linewidth]{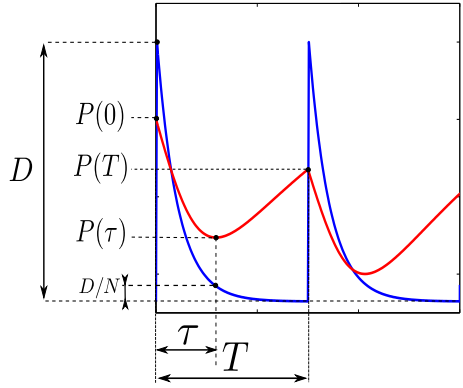}
\caption{\textbf{The action of chemotherapy}. A cycle of chemotherapy of length $T$ can be subdivided into two time intervals. The red curve represents the time series of the tumor size $P(t)$, while the drug concentration $C(t)$ is added for clarity (blue curve). During the first interval $[0,\tau)$, when the drug concentration is high $C(t)>D/N$, the chemotherapeutic drugs govern the dynamics and the tumor is reduced from its original size $P(0)$ to a size $P(\tau)$. During the second interval $[\tau, T)$, which starts when the drug concentration has dropped to low levels, the tumor regrows from $P(\tau)$ to its final size at the end of the cycle $P(T)$. Note that, by definition, the inequality $\tau \leq T$ always holds.}
\label{fig:4}
\end{figure}

To obtain a function that relates the threshold value of $T$ with the other parameters of the model, we first derive a one dimensional map for cancer chemotherapy. In dynamical systems, a map is a function that relates a set of possible states with itself in an iterative manner. In our case, this function relates the size of the tumor right before two successive cycles of chemotherapy. The first assumption in this derivation is that the drugs are effective enough to reduce the size of the tumor at some time. An upper bound for the value of $r$ that guarantees this condition is $r<b(1-e^{-\rho D})$, which can be obtained by considering that $r P(1-P/K) \leq r P<b \left(1-e^{-\rho C(t)} \right)P$ in equation \eqref{eq:2}, which assures that $\dot{P}<0$. Rearranging this condition we can also set a restriction on the minimal dose of drug required for a treatment to be effective, which is $\rho D > \log_{e}(b/(b-r))$. For this not to hold, the particular nature of the protocol would not have much importance, since the only effect of chemotherapy is just to slow the growth of the tumor, but not to reduce it. As depicted in Fig.~\ref{fig:4}, when the drug concentration is high, the second term on the right hand side of equation \eqref{eq:2} dominates over the first term. Conversely, once most of the drug has been eliminated $C(t)\rightarrow 0$, it occurs the other way around, \emph{i.e.}, the second term is vanishingly small in comparison to the first. Thus, we can divide a cycle of chemotherapy $[0,~T)$ into two time intervals. The first interval $[0,~\tau)$ is dominated by the cytotoxic drugs, and during it, equation \eqref{eq:2} can be approximated as
\begin{equation}
\dfrac{d P}{d t}=-b \left(1-e^{-\rho C(t)} \right) P(t). 
\label{eq:7}
\end{equation}
During the second interval $[\tau,~T)$ the tumor growth prevails, and therefore we can consider the growth term only
\begin{equation}
\dfrac{d P}{d t}=r P(t) \left(1-\dfrac{P(t)}{K}\right).
\label{eq:8}
\end{equation}

These two equations can be easily integrated. However, a difficulty arises in order to estimate $\tau$, which represents the time at which the drug concentration has decreased to values for which the first term on the right hand side of \eqref{eq:2} starts to dominate over the second. One possibility is to approximate the growth of the tumor to exponential. Since this value is the maximum rate at which the tumor can grow (occurring for small tumor burdens), the value thus obtained is clearly a conservative overestimation. In this case, making use of equation \eqref{eq:6}, the following inequality must hold
\begin{equation}
r \geq b \left(1-e^{-\rho D e^{-k t}} \right).
\label{eq:9}
\end{equation}
The value at saturation can be used to solve for $\tau$, yielding
\begin{equation}
\tau = \dfrac{1}{k}\log_{e} \left(\rho D\small/\log_{e} \left(b/(b-r)\right) \right).
\label{eq:10}
\end{equation}

A simpler and less restrictive possibility is to assume that when the drug concentration has dropped to a certain level, its cytotoxic effect is negligible. Following this reasoning, we can write $\tau=(\log_{e}N)/k$, where $N$ represents the fraction to which the drug concentration has dropped. Of course, comparison of this equation with equation \eqref{eq:10} allows to solve for $N$. More generally, a prudent choice could be to consider that when the concentration has reduced in an order of magnitude ($N=10$), the effects of the drug can be disregarded. In fact, for the parameter values presented in the previous section and this criterion, the estimated values of the time $\tau$ obtained through these two different methods are quite similar (around four days). In what follows, for simplicity, and because the conclusions are more resounding, we consider a fixed value of $N$ to illustrate our results. Nevertheless, the first approach can be used to obtain similar conclusions, because $\tau$ grows slowly when $\rho D$ is increased. 

Having explained this point, we can proceed to integrate the equations \eqref{eq:7} and \eqref{eq:8}. Equation \eqref{eq:7} can be integrated as follows. In a first step we have
\begin{equation}
\log_{e}(P(\tau)/P(0)) = -b \tau + b \int_{0}^{\tau} e^{-\rho D e^{-k t}} dt.
\label{eq:11}
\end{equation}
The integral appearing in the second term of the right hand side can be solved in terms of the exponential integral function $\text{Ei}(x)$ (see Appendix \ref{app}). The result is
\begin{equation}
\int_{0}^{\tau} e^{-\rho D e^{-k t}} dt = \tau  \dfrac{\text{Ei}(-\rho D)-\text{Ei}(-\rho D/N)}{\log_{e}N}.
\label{eq:12}
\end{equation}
If we define the fuction $g$ as
\begin{equation}
g(\rho D,N) \equiv \dfrac{\text{Ei}(-\rho D)-\text{Ei}(-\rho D/N)}{\log_{e}N},
\label{eq:13}
\end{equation}
we finally obtain
\begin{equation}
P(\tau)=P(0)e^{-b \tau(1-g)}.
\label{eq:14}
\end{equation}

\begin{figure}
\centering
  \includegraphics[width=0.44\linewidth,height=0.46\linewidth]{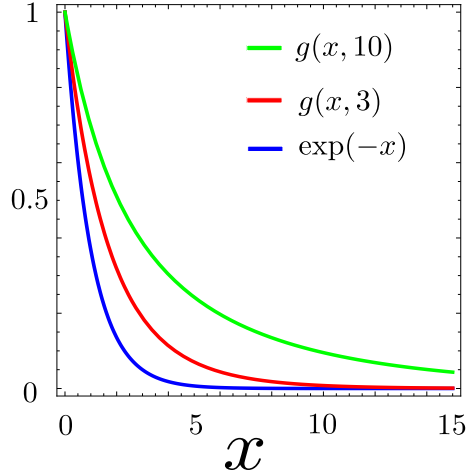}
\caption{\textbf{The function $\boldsymbol{g(x,N)}$}. The resemblance between this function and a decaying exponential is shown. For $x=0$ it takes a value of $1$ and it has an horizontal asymptote at $y=0$.}
\label{fig:5}
\end{figure}

The function $g(x,N)$ is analytic in the domain of interest $[0,\infty) \times (1,\infty)$ and resembles an exponential decay, as can be seen in Fig.~\ref{fig:5}. This means that the survival fraction $P(\tau)/P(0)$ at time $\tau$ plateaus for increasing values of the dose. The value of the survival fraction at the plateau is $e^{-b \tau}$. This feature is characteristic of the Exponential Kill Model and does not appear in ordinary log-kill models \citep{gard}. It states that at a certain point, which depends on the tumor resistance to the drugs, the increase of the dose intensity barely affects the survival fraction.

The solution to the logistic equation \eqref{eq:8} governing the second part of the chemotherapeutic cycle $[\tau, T)$ is well-known. With these limits of integration, the solution can be written as
\begin{equation}
P(T)=\dfrac{K P(\tau) e^{r (T-\tau)}}{K+P(\tau)(e^{r(T-\tau)}-1)}.
\label{eq:15}
\end{equation}

To conclude, we can substitute equation \eqref{eq:14} into equation \eqref{eq:15} to obtain the relation between the size of the tumor at the beginning and at the end of a cycle of chemotherapy
\begin{equation}
P(T)=\dfrac{K P(0)}{K e^{-r(T-\tau(1+b(1-g)/r))}+P(0)(1-e^{-r(T-\tau)})}.
\label{eq:16}
\end{equation}
If we define the constants $\alpha=e^{-r(T-\tau(1+b(1-g)/r))}$ and $\beta=1-e^{-r(T-\tau)}$, the sequence that relates the size of the tumor cell population right when the $n$-th cycle starts and its size at the end of the cycle is
\begin{equation}
P_{n+1}=\dfrac{K P_{n}}{\alpha K+\beta P_{n}}.
\label{eq:17}
\end{equation}
In fact, we can nondimensionalize the tumor size population by dividing it by its carrying capacity, since it will not affect the results presented in the following section. In the new variable $x=P/K$, the map is nicely written as
\begin{equation}
x_{n+1}=f(x_{n})=\dfrac{x_{n}}{\alpha+\beta x_{n}}.
\label{eq:18}
\end{equation}

\section{The shrinking condition}\label{sec:sco}

We proceed to study the stability properties of the one-dimensional map derived right above. The fixed points of the map satisfy the equation $f(x^{*})=x^{*}$. If chemotherapy is able to completely reduce the size of the tumor, the orbit of any initial condition should asymptotically converge to a state of zero tumor cell population. In mathematical language, $x^{*}=0$ should be an attractor of the dynamical system. This imposes a constraint on the values that $\alpha$ can take. Solving for $x^{*}$, we find two possible solutions, which are $x_{1}^{*}=0$ and $x_{2}^{*}=(1-\alpha)/\beta$. To address the stability of the fixed points, the Jacobian has to be computed at these points. We obtain
\begin{equation}
f'(x^{*})=\dfrac{\alpha}{(\alpha+\beta x^{*})^2}.
\label{eq:19}
\end{equation}
Thus, for $x_{1}^{*}=0$, the value of the Jacobian is $1/\alpha$, while for $x_{2}^{*}=(1-\alpha)/\beta$, the value is $\alpha$. Since $\alpha$ is positive, we have two possibilities. The first one is $\alpha<1$, where $x_{1}$ is a repelling fixed point, while $x_{2}$ represents a stable state in which the tumor exists under its carrying capacity. Conversely, if $\alpha>1$, the tumor can be eradicated, which is the desired situation. We can solve for $T$ by expressing the parameter $\alpha$ in terms of the parameters of the chemotherapeutic protocol and the rate of growth of the tumor. The result is
\begin{equation}
T<\tau(1+b(1-g(\rho D,N))/r).
\label{eq:20}
\end{equation}

As it can be seen, there certainly exists a threshold value of the period between cycles $T_{s}=\tau(1+b(1-g(\rho D,N))/r)$ for a chemotherapeutic protocol to be effective. On what follows, we refer to this value as the \emph{shrinking time} $T_{s}$. Protocols that are sufficiently dose-dense (\emph{i.e., $T<T_{s}$}) provide a sustained reduction of the tumor. Note that, since our estimation of the time $\tau$ does not depend on $K$, there is no dependence of $T_{s}$ on the carrying capacity of the tumor $K$. This is an advantage, since, in general, it is not easy to determine \emph{a priori} the value of $K$.
\begin{figure}
\centering
  \includegraphics[width=0.52\linewidth,height=0.46\linewidth]{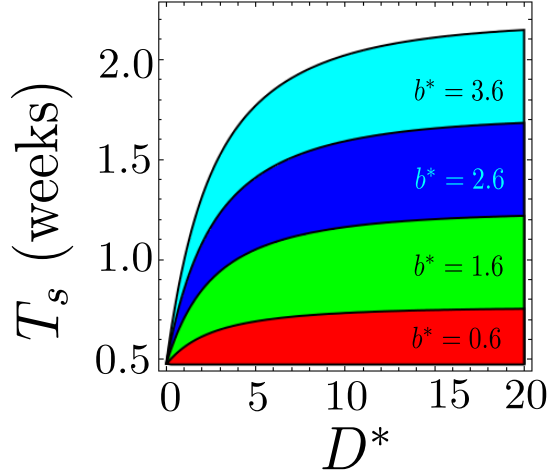}
\caption{\textbf{The shrinking condition $\boldsymbol{T_{s}}$}. The curves $T_{s}$ are represented against the effective dose $D^{*}$ and for different values of the relative maximum fractional cell kill $b^{*}$. The colored regions $T<T_{s}$ represent sufficiently dose-dense protocols. Note how the curves asymptote to a value $T_{\infty}$, showing that the increase in dose does not provide substantial benefits if the density of the protocol is not enough.}
\label{fig:6}
\end{figure}

We now examine the nature of the functional relation between $T_{s}$ and the parameters representing the dose $D$, the resistance $\rho$, the maximum fractional cell kill $b$ and the rate of growth $r$ of the tumor. Since these four parameters appear in pairs in equation \eqref{eq:20}, $\rho$ and $D$ multiplying each other while $b$ and $r$ dividing each other, we can simplify the study by defining two new dimensionless parameters from them. We call one of them the effective dose $D^{*}=\rho D$, while the other is named the relative maximum fractional cell kill $b^{*}=b/r$. In the following we consider $N=10$, which, as previously stated, gives similar results that equation \eqref{eq:10}. With the new parameters, the shrinking time can be written as $T_{s}=(\log_{e}10)(1+b^{*}(1-g(D^{*},10)))/k$, where we use the same value of $k$ as before. As it is shown in Fig.~\ref{fig:6}, the higher the effective dose $D^{*}$, the higher the shrinking time can be raised. However, it is clearly appreciated that the value plateaus for high doses, and a point is reached for which increasing the dose intensity does not allow to reduce noticeably the dose density. This is an important prediction of the present work. Increasing the dose of a protocol might not be very useful (even if there was no toxicity) if its density is under a certain value. An approximation of the limiting value is attained for $D^{*} \rightarrow \infty$ and can be computed as $T_{\infty}=\tau(1+b^{*})$. As shown in Fig.~\ref{fig:6}, typical values of this time span from less than a week to several weeks. Concerning the relative fractional cell kill, it is evident that the shrinking time increases with it. Thus, in our model, slower growing tumors and higher maximum fractional cell kill (more destructive drugs) permit to reduce the density of the protocol. 

Even though the existence of a shrinking time does not depend on additional hypothesis, the precise value that it takes, certainly does. It is at this point that the Norton-Simon hypothesis comes into play. For tumors that have less proliferating cells, the effects of chemotherapy are expected to be smaller, since chemotherapy is more effective on proliferating than quiescent cells. This phenomenon would maintain a similar value of $b^{*}$, by reducing the value of $b$. Therefore, our results are also valid for other types of tumor growth, and are always relevant as long as the relative maximum fractional cell kill is sufficiently small. For example, for haematological cancers we can use exponential growth $\dot{P}=r P$ instead of equation \eqref{eq:8} and derive equivalent equations. There is also evidence pointing to the fact that some solid tumors do not follow a sigmoidal growth, and that their mean radius increases linearly with time \citep{bru}. This occurs because only the cells on the surface of these tumors are proliferating. Again, shrinking times can be computed following our recipe, with $\dot{P}=r P^{2/3}$. In principle, a tumor that follows this power law tends to grow considerably more slowly than an exponentially growing tumor with the same constant rate, because of the factor $2/3$ in the exponent. In the case of a single connected tumor for which only cells on its surface grow, the constant rate is similar to the exponentially growing case. Its value can be computed as $r=(36 \cdot \pi \rho)^{1/3}c$, where $\rho$ is the volumetric density of cells and $c$ is the rate at which the radius grows. Even if the tumor is formed by a number $N$ of pieces, as for example a disconnected tumor or a tumor that has metastasized, the constant rate can be estimated as $r=(36 \cdot \pi \rho N)^{1/3}c$, which is similar to the growth constant rates here considered. However, this does not mean that the value $b^{*}$ is necessarily higher, because the maximum fractional cell kill of chemotherapy $b$, according to the Norton-Simon hypothesis, decreases as well.

\section{Combination protocols}\label{sec:comb}

\begin{figure}
\centering
  \includegraphics[width=0.9\linewidth,height=0.3\linewidth]{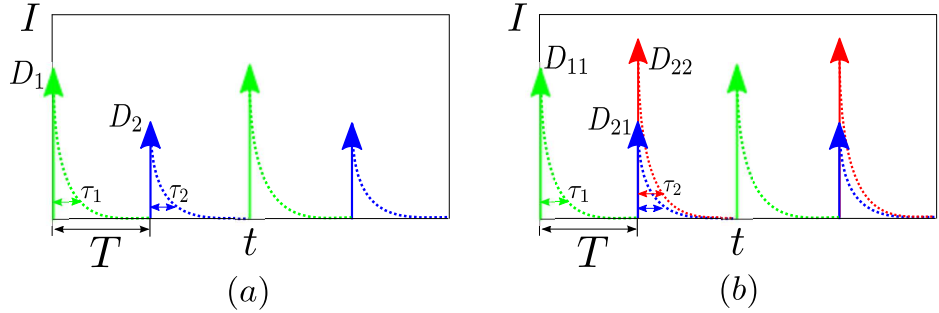}
\caption{\textbf{Protocols with several drugs}. (a) An alternate protocol consisting of two drugs. (b) A protocol consisting of two simultaneously administered drugs alternated with a single administered drug. Note how in the latter case the time $\tau$ is approximated to the highest value of the two drugs (red and blue).}
\label{fig:7}
\end{figure}

Generally, several drugs are combined in a protocol of chemotherapy. For example, an ordinary protocol of chemotherapy for locally advanced breast cancer can combine at least three cytotoxic drugs, such as cyclophosphamide, epirubicin and flurouracil \citep{drugs}. It is therefore pertinent to ask if the map obtained in Sec.~\ref{sec:1dm} can be extended to derive the shrinking conditions for more complex protocols. Fortunately, the answer is affirmative. We first consider an imaginary protocol consisting of two drugs (see Fig.~\ref{fig:7}), which are given in an alternate fashion (a cycle of the first drug followed by a cycle of the other drug). Since each drug has its own parameters $\alpha_{i}$ and $\beta_{i}$, after the first cycle we have
\begin{equation}
x_{1}=f_{1}(x_{0})=\dfrac{x_{0}}{\alpha_{1}+\beta_{1} x_{0}}.
\label{eq:21}
\end{equation}
Then, the second cycle is applied
\begin{equation}
x_{2}=f_{2}(x_{1})=\dfrac{x_{1}}{\alpha_{2}+\beta_{2} x_{1}}.
\label{eq:22}
\end{equation}
Now, we have to note that equation \eqref{eq:18} can be regarded as a M\"obius transformation, restricted to the real numbers and with two parameters fixed. This means that the composition of the two maps yields the same map, but with two different parameters $(\alpha,\beta)$. Mathematically we have $f=f_{2} \circ f_{1}$, with $x_{2}=f(x_{0})$. The new parameters are related to the old ones through the relations
\begin{equation}
\alpha=\alpha_{1}\alpha_{2}, \beta=\beta_{2}+\alpha_{2}\beta_{1},
\label{eq:23}
\end{equation}
which define a Lie group relation between the group elements $g=(\alpha,\beta)$ and can be represented as the product of matrices $R(g)$ of the form
\begin{equation}
R(g)=\left( {\begin{array}{cc}\alpha & 0 \\       \beta & 1 \      \end{array} } \right).
\label{eq:24}
\end{equation}
In other words, there is an underlying Lie group structure that allows us to reduce the complex protocol to a simple one. This group representation clearly resembles to the affine group. In fact, if $\alpha$ and $\beta$ belonged to $\mathbb{R}_{>0}$ and $\mathbb{R}$ respectively, $f$ would define an action of the half-plane on the real line. However, in our case $\alpha$ and $\beta$ belong to the interval $[0,1]$, and therefore, the inverse element is lost (there is no such thing as a cycle of antichemotherapy). In summary, we can combine alternated drugs in a protocol with no difficulty to give estimations of the shrinking time. Since we have $\alpha>1$, and $\alpha_{1}$ and $\alpha_{2}$ are exponentials, for the present example we have
\begin{equation}
T_{s}=\dfrac{1}{2}(\tau_{1} (1+b_{1}(1-g(\rho_{1} D_{1},N))/r)+\tau_{2}(1+b_{2}(1-g(\rho_{2} D_{2},N))/r)).
\label{eq:25}
\end{equation}
Thus, the new shrinking time is the arithmetic mean of the shrinking times of both drugs. Finally, two or more drugs are frequently given simultaneously. The general differential equation for a protocol with $n_{d}$ non-interacting drugs given simultaneously, can be written as
\begin{equation}
\dfrac{d P}{d t}=r P(t) \left(1-\dfrac{P(t)}{K} \right)-\sum_{j=0}^{n_{d}}b_{j} \left(1-e^{-\rho_{j} C_{j}(t)} \right) P(t).
\label{eq:26}
\end{equation}
In this case, if the drugs have very different half-lives, we can derive again the same map, but the development is more complicated. Note that, to estimate $\tau$ to construct the map, we have to choose between the different $\tau_{j}$ of each drug ($j=1,...,n_{d}$). We can consider as an approximation the maximum of these times $\max \tau_{j}$, but this procedure complicates the integrals associated to the other drugs, and their solution involves the gamma function. However, if the half-lives of the drugs are not so different, we can approximate all the $\tau_{j}$ to a single value. Then, the solution is much more simple. For example, with two drugs given simultaneously we have the shrinking time 
\begin{equation}
T_{s}=\tau_{1} (1+b_{11}(1-g(\rho_{11} D_{11},N))/r+b_{12}(1-g(\rho_{12} D_{12},N))/r),
\label{eq:27}
\end{equation}
where a matrix notation has been adopted for the parameters $b$, $\rho$ and $D$, corresponding to each chemotherapeutic drug. This equation tells us that giving more drugs at a time allows us to relax the frequency (dose-density) of the protocol. Of course, this comes at the expense of more toxicity.

Therefore, a general approximation of the shrinking time for an arbitrary protocol, as shown in Fig.~\ref{fig:7}, as long as the half-lives of simultaneously given drugs are similar, is
\begin{equation}
T_{s}=\dfrac{1}{n}\sum_{i=1}^{n}\sum_{j=1}^{n_{i}}\tau_{i} (\delta_{ij}+b_{ij}(1-g(\rho_{ij} D_{ij},N))/r),
\label{eq:28}
\end{equation}
where $\delta_{ij}$ is the Kronecker delta and $\tau_{i}$ is the time for which simultaneously given drugs have decayed to sufficiently low levels. The variable $n_{i}$ is the number of simultaneous drugs given at the $i$-th step of an alternate protocol comprising $n$ steps. To conclude this section, we recall that if two different drugs are given sequentially (some cycles of the first followed by a number of cycles of the second), the whole treatment can be reduced to two subtreatments, having its own shrinking time value each.

\section{Conclusions}\label{sec:con}

The existence of a shrinking time is a sound argument that supports the use of dose-dense protocols and it is in conformity with other works on chemotherapy \citep{dd4,norton}, which defend the equal importance of dose-density and dose-intensity. However, even though the administration of two-week cycles of chemotherapy represents statistically significant advantages compared to the conventional three-week administration, these benefits are modest \citep{dd4}. Moreover, recent investigations demonstrate that recurrence-free survival rates are not substantially improved by using tailored dose-dense chemotherapy instead of standard chemotherapy over a median of five years \citep{notwo}. Among other reasons, this could be explained by considering that, although dose-dense protocols can reduce significantly the survival fraction at the nadir, this reduction is not enough to avoid the regrowth of residual tumor cells. In addition, it must be recognized that, at some point, the introduction of more dose-dense protocols presents similar difficulties that the increase of dose-intensity. The increase of dose-density achieved by reducing the cycles below one week would be a synonym of an increase in dose-intensity. Clearly put, in the limit of small times between cycles, dose-density is tantamount to dose escalation, which can introduce intolerable toxicities. Finally, we recall that part of the cells that comprise a solid tumor are frequently found in a quiescent state. In such a case too dose-dense protocols might not work, since quiescent cells are less susceptible to chemotherapy and some time after a cycle has caused its destruction might be necessary for these remaining cells to abandon their cell cycle arrest. Therefore, there is no doubt that dose-density presents numerous difficulties.

As our study suggests, chemotherapy might sometimes present the following contradiction. As time is given for the side-effects of chemotherapy to disappear and for the organism to restore its homeostasis, time is also given for the tumor to recuperate. In light of the facts argued in the previous paragraph, it is worth to examine if other treatment strategies, in addition to dose-densification, are possible. As an example, we wonder if some kind of targeted cytostatic drugs (whenever they exist) can be administered in a rather continuous fashion between cycles of chemotherapy to arrest the growth of the tumor during that time. Were this possible, the targeted cytostatic drug would have the effect of arresting the growth of the tumor only, while allowing the regrowth of healthy cells that are also destroyed by the cytotoxic drugs. We acknowledge that this method presents other difficulties, because it requires some degree of synchronization. If the cytostatic effect of the drug persists for the time a new cycle of cytotoxic drugs starts, the effectiveness of the last can be reduced. Again, this is so because chemotherapy is more efficient on dividing than quiescent cells. Moreover, if we could arrest the growth of tumor cells by means of some cytostatic targeted drug, the periodicity of the cycles might be relaxed and the toxic effects of chemotherapeutic drugs would be reduced. This, in turn, would perhaps allow an increase in the number of cycles as well.

In summary, we wonder if therapies might be improved as well if non-selective destructive drugs are alternated with more specific and targeted drugs, which can prevent the regrowth of the tumor cells between cycles. Taking up the old concept of the magic bullet \citep{mbul} proposed by Paul Ehrlich to denominate these selective drugs, we suggest that the cannonballs of traditional cytotoxic chemotherapy might be complemented with the magic bullets of some cytostatic targeted therapy administered between cycles. In principle, the idea is fairly simple and testable. Between cycles of destruction, stasis.

\section*{Aknowledgments}  \label{sec:aknow}

This work has been supported by the Spanish Ministry of Economy and Competitiveness under Projects No. FIS2013-40653-P and by the Spanish State Research Agency (AEI) and the European Regional Development Fund (FEDER) under Project No. FIS2016-76883-P. M.A.F.S. acknowledges the jointly sponsored financial support by the Fulbright Program and the Spanish Ministry of Education (Program No. FMECD-ST-2016). A.G.L. akcnowledges the financial support of \emph{Programa de Becas Iberoam\'{e}rica. Santander Universidades. Espa\~{n}a, 2015} for a research stay at the Universidade Federal do Paran\'{a}. K.C.I. acknowledges FAPESP (2015/07311-7).

\begin{appendices}
\renewcommand{\theequation}{A\arabic{equation}}
\setcounter{equation}{0} 

\section{Solution to the integral} \label{app}
 
In Sec. \ref{sec:1dm}, the following result has been used
\begin{equation}
\int_{0}^{\tau} e^{-D^{*} e^{-k t}} dt = \tau  \dfrac{\text{Ei}(-D^{*})-\text{Ei}(-D^{*}/N)}{\log_{e}N},
\label{eq:app1}
\end{equation}
where $\text{Ei}(x)$ is the exponential integral function, which is defined as
\begin{equation}
\text{Ei}(x)=\int_{-\infty}^{x} \dfrac{e^{t}}{t} dt.
\label{eq:app2}
\end{equation}

Here we demonstrate the equality \eqref{eq:app1}. To this end, we first perform a change of variables $u=e^{-k t}$. Therefore, the relation between the differential elements is $d u/u=-k d t$. Recalling that we have considered $\tau=(\log_{e}N)/k$, the integral in the new variable reads
\begin{equation}
\int_{0}^{\tau} e^{-D^{*} e^{-k t}} dt =-\dfrac{\tau}{\log_{e}N} \int_{1}^{1/N} \dfrac{e^{-D^{*} u}}{u} du.
\label{eq:app3}
\end{equation}
Now, performing another change of variables $x=-D^{*}u$, we can rewrite the previous integral as
\begin{equation}
\int_{1}^{1/N} \dfrac{e^{-D^{*} u}}{u} du=\int_{-D^{*}}^{-D^{*}/N} \dfrac{e^{x}}{x} dx.
\label{eq:app4}
\end{equation}
Finally, the last integral can be expressed as the sum of two integrals in the following way
\begin{equation}
\int_{-D^{*}}^{-D^{*}/N} \dfrac{e^{x}}{x} dx=\int_{-\infty}^{-D^{*}/N} \dfrac{e^{x}}{x} dx-\int_{-\infty}^{-D^{*}} \dfrac{e^{x}}{x} dx.
\label{eq:app5}
\end{equation}
In this last step, the conditions $D^{*}>0$ and $N>1$ have been used, since the $\text{Ei}(x)$ has a singularity at $x=0$. The integrals appearing in this last equation are precisely the $\text{Ei}(x)$ function.

\end{appendices}

\end{document}